\pdfoutput=1

\documentclass[12pt,a4paper]{article}

\usepackage{ifthen} 
\newboolean{pdflatex}
\setboolean{pdflatex}{true} 

\newboolean{articletitles}
\setboolean{articletitles}{true} 

\newboolean{uprightparticles}
\setboolean{uprightparticles}{false} 

\newboolean{inbibliography}
\setboolean{inbibliography}{false} 

\usepackage{microtype}
\usepackage{lineno}  
\usepackage{xspace} 
\usepackage{caption} 

\usepackage{graphicx}  
\usepackage{color}
\usepackage{colortbl}
\graphicspath{{./figs/}} 

\usepackage{amsmath} 
\usepackage{amssymb}
\usepackage{amsfonts}
\usepackage{upgreek} 

\newcommand*\patchAmsMathEnvironmentForLineno[1]{%
\expandafter\let\csname old#1\expandafter\endcsname\csname #1\endcsname
\expandafter\let\csname oldend#1\expandafter\endcsname\csname
end#1\endcsname
 \renewenvironment{#1}%
   {\linenomath\csname old#1\endcsname}%
   {\csname oldend#1\endcsname\endlinenomath}%
}
\newcommand*\patchBothAmsMathEnvironmentsForLineno[1]{%
  \patchAmsMathEnvironmentForLineno{#1}%
  \patchAmsMathEnvironmentForLineno{#1*}%
}
\AtBeginDocument{%
\patchBothAmsMathEnvironmentsForLineno{equation}%
\patchBothAmsMathEnvironmentsForLineno{align}%
\patchBothAmsMathEnvironmentsForLineno{flalign}%
\patchBothAmsMathEnvironmentsForLineno{alignat}%
\patchBothAmsMathEnvironmentsForLineno{gather}%
\patchBothAmsMathEnvironmentsForLineno{multline}%
\patchBothAmsMathEnvironmentsForLineno{eqnarray}%
}

\def\lhcb {\mbox{LHCb}\xspace}

\ifthenelse{\boolean{uprightparticles}}%
{

 \def\Peta        {\ensuremath{\upeta}\xspace}

 \def\Ppi         {\ensuremath{\uppi}\xspace}

 \def\Ppsi        {\ensuremath{\uppsi}\xspace}

 \def\PDelta      {\ensuremath{\Delta}\xspace}                 
 \def\PXi      {\ensuremath{\Xi}\xspace}                 
 \def\PLambda      {\ensuremath{\Lambda}\xspace}                 
 \def\PSigma      {\ensuremath{\Sigma}\xspace}                 
 \def\POmega      {\ensuremath{\Omega}\xspace}                 
 \def\PUpsilon      {\ensuremath{\Upsilon}\xspace}                 
 
 \def\PB      {\ensuremath{\mathrm{B}}\xspace}                 
                  
 \def\PD      {\ensuremath{\mathrm{D}}\xspace}

 \def\PJ      {\ensuremath{\mathrm{J}}\xspace}                 
 \def\PK      {\ensuremath{\mathrm{K}}\xspace}

 \def\Pc      {\ensuremath{\mathrm{c}}\xspace}

 \def\Pi      {\ensuremath{\mathrm{i}}\xspace}

 \def\Pp      {\ensuremath{\mathrm{p}}\xspace}

}
{

 \def\Peta        {\ensuremath{\eta}\xspace}

 \def\Ppi         {\ensuremath{\pi}\xspace}

 \def\Ppsi        {\ensuremath{\psi}\xspace}                 
                  
 \mathchardef\PDelta="7101
 \mathchardef\PXi="7104
 \mathchardef\PLambda="7103
 \mathchardef\PSigma="7106
 \mathchardef\POmega="710A
 \mathchardef\PUpsilon="7107
                  
 \def\PB      {\ensuremath{B}\xspace}                 
                  
 \def\PD      {\ensuremath{D}\xspace}

 \def\PJ      {\ensuremath{J}\xspace}                 
 \def\PK      {\ensuremath{K}\xspace}

 \def\Pc      {\ensuremath{c}\xspace}

 \def\Pi      {\ensuremath{i}\xspace}

 \def\Pp      {\ensuremath{p}\xspace}

}

\makeatletter
\ifcase \@ptsize \relax
  \newcommand{\miniscule}{\@setfontsize\miniscule{4}{5}}
\or
  \newcommand{\miniscule}{\@setfontsize\miniscule{5}{6}}
\or
  \newcommand{\miniscule}{\@setfontsize\miniscule{5}{6}}
\fi
\makeatother

\DeclareRobustCommand{\optbar}[1]{\shortstack{{\miniscule (\rule[.5ex]{1.25em}{.18mm})}
  \\ [-.7ex] $#1$}}

\def\cquark    {{\ensuremath{\Pc}}\xspace}

\def\pion   {{\ensuremath{\Ppi}}\xspace}

\def\pip    {{\ensuremath{\pion^+}}\xspace}
\def\pim    {{\ensuremath{\pion^-}}\xspace}
\def\pipm   {{\ensuremath{\pion^\pm}}\xspace}

\def\kaon    {{\ensuremath{\PK}}\xspace}
  \def\Kbar    {{\kern 0.2em\overline{\kern -0.2em \PK}{}}\xspace}

\def\KorKbar    {\kern 0.18em\optbar{\kern -0.18em K}{}\xspace}

\def\Kp      {{\ensuremath{\kaon^+}}\xspace}
\def\Km      {{\ensuremath{\kaon^-}}\xspace}
\def\Kpm     {{\ensuremath{\kaon^\pm}}\xspace}

  \def\Dbar    {{\kern 0.2em\overline{\kern -0.2em \PD}{}}\xspace}

\def\DorDbar    {\kern 0.18em\optbar{\kern -0.18em D}{}\xspace}

\def\B       {{\ensuremath{\PB}}\xspace}
\def\Bbar    {{\ensuremath{\kern 0.18em\overline{\kern -0.18em \PB}{}}}\xspace}

\def\BorBbar    {\kern 0.18em\optbar{\kern -0.18em B}{}\xspace}

\def\Bu      {{\ensuremath{\B^+}}\xspace}
\def\Bub     {{\ensuremath{\B^-}}\xspace}
\def\Bp      {{\ensuremath{\Bu}}\xspace}
\def\Bm      {{\ensuremath{\Bub}}\xspace}
\def\Bpm     {{\ensuremath{\B^\pm}}\xspace}

\def\jpsi     {{\ensuremath{{\PJ\mskip -3mu/\mskip -2mu\Ppsi\mskip 2mu}}}\xspace}
\def\psitwos  {{\ensuremath{\Ppsi{(2S)}}}\xspace}

\def\etac     {{\ensuremath{\Peta_\cquark}}\xspace}

  \def\Y#1S{\ensuremath{\PUpsilon{(#1S)}}\xspace}

\def\proton      {{\ensuremath{\Pp}}\xspace}
\def\antiproton  {{\ensuremath{\overline \proton}}\xspace}

\def\Lz          {{\ensuremath{\PLambda}}\xspace}
\def\Lbar        {{\ensuremath{\kern 0.1em\overline{\kern -0.1em\PLambda}}}\xspace}
\def\LorLbar    {\kern 0.18em\optbar{\kern -0.18em \PLambda}{}\xspace}

\def\to                 {\ensuremath{\rightarrow}\xspace}

\def\CP                {{\ensuremath{C\!P}}\xspace}

\def\AFB      {\ensuremath{A_{\mathrm{FB}}}\xspace}

\def\AT#1     {\ensuremath{A_{\mathrm{T}}^{#1}}\xspace}           

\def\C#1      {\ensuremath{\mathcal{C}_{#1}}\xspace}                       
\def\Cp#1     {\ensuremath{\mathcal{C}_{#1}^{'}}\xspace}                    
\def\Ceff#1   {\ensuremath{\mathcal{C}_{#1}^{\mathrm{(eff)}}}\xspace}        
\def\Cpeff#1  {\ensuremath{\mathcal{C}_{#1}^{'\mathrm{(eff)}}}\xspace}       
\def\Ope#1    {\ensuremath{\mathcal{O}_{#1}}\xspace}                       
\def\Opep#1   {\ensuremath{\mathcal{O}_{#1}^{'}}\xspace}                    

\newcommand{\tev}{\ifthenelse{\boolean{inbibliography}}{\ensuremath{~T\kern -0.05em eV}\xspace}{\ensuremath{\mathrm{\,Te\kern -0.1em V}}}\xspace}
\newcommand{\gev}{\ensuremath{\mathrm{\,Ge\kern -0.1em V}}\xspace}
\newcommand{\mev}{\ensuremath{\mathrm{\,Me\kern -0.1em V}}\xspace}
\newcommand{\kev}{\ensuremath{\mathrm{\,ke\kern -0.1em V}}\xspace}
\newcommand{\ev}{\ensuremath{\mathrm{\,e\kern -0.1em V}}\xspace}
\newcommand{\gevc}{\ensuremath{{\mathrm{\,Ge\kern -0.1em V\!/}c}}\xspace}
\newcommand{\mevc}{\ensuremath{{\mathrm{\,Me\kern -0.1em V\!/}c}}\xspace}
\newcommand{\gevcc}{\ensuremath{{\mathrm{\,Ge\kern -0.1em V\!/}c^2}}\xspace}
\newcommand{\gevgevcccc}{\ensuremath{{\mathrm{\,Ge\kern -0.1em V^2\!/}c^4}}\xspace}
\newcommand{\mevcc}{\ensuremath{{\mathrm{\,Me\kern -0.1em V\!/}c^2}}\xspace}

\def\invfb   {\ensuremath{\mbox{\,fb}^{-1}}\xspace}

\def\gsim{{~\raise.15em\hbox{$>$}\kern-.85em
          \lower.35em\hbox{$\sim$}~}\xspace}
\def\lsim{{~\raise.15em\hbox{$<$}\kern-.85em
          \lower.35em\hbox{$\sim$}~}\xspace}

\def\evtgen     {\mbox{\textsc{EvtGen}}\xspace}

\def\gauss      {\mbox{\textsc{Gauss}}\xspace}
\def\geant      {\mbox{\textsc{Geant4}}\xspace}

\def\photos     {\mbox{\textsc{Photos}}\xspace}

\def\pythia     {\mbox{\textsc{Pythia}}\xspace}

\def\tell1  {TELL1\xspace}
\def\ukl1   {UKL1\xspace}

\usepackage{cite} 
\usepackage{mciteplus}
\usepackage{hyperref}    
\usepackage[all]{hypcap} 

\def\hhh {\ensuremath{\Bpm \to h^{\pm} h^+ h^-}\xspace}

\def\pppi {\ensuremath{\Bpm \to p \antiproton \pi^{\pm}}\xspace}
\def\ppk {\ensuremath{\Bpm \to p \antiproton K^{\pm}}\xspace}

\def\pipipiplus {\ensuremath{B^+ \to \pip \pim \pip}\xspace}
\def\kpipiplus {\ensuremath{B^+ \to \Kp \pim \pip}\xspace}
\def\kkpiplus {\ensuremath{B^+ \to \Kp \Km \pip}\xspace}
\def\kkkplus {\ensuremath{B^+ \to \Kp \Km \Kp}\xspace}
\def\pppiplus {\ensuremath{B^+ \to p \antiproton  \pip}\xspace}
\def\ppkplus {\ensuremath{B^+ \to p \antiproton  \Kp}\xspace}
\def\pphplus {\ensuremath{B^+ \to p \antiproton  h^+}\xspace}

\def\lambdaFifteenTwentypplus {\ensuremath{B^+ \to \Lbar(1520)p}\xspace}
\def\lambdaFifteenTwentypplusdet {\ensuremath{B^+ \to \Lbar(1520)(\to \Kp\antiproton)p}\xspace}

\def\acp {\ensuremath{A_{\CP}}\xspace}
\def\acprawacc {\ensuremath{A_{\rm raw}^{\rm{acc}}}\xspace}
\def\acpraw {\ensuremath{A_{\rm raw}}\xspace}

\def\aprod {\ensuremath{A_{\rm P}}\xspace}
\def\adet {\ensuremath{A_{\rm det}}\xspace}

\begin{document}

\renewcommand{\thefootnote}{\fnsymbol{footnote}}
\setcounter{footnote}{1}


\begin{titlepage}
\pagenumbering{roman}

\vspace*{-1.5cm}
\centerline{\large EUROPEAN ORGANIZATION FOR NUCLEAR RESEARCH (CERN)}
\vspace*{1.5cm}
\hspace*{-0.5cm}
\begin{tabular*}{\linewidth}{lc@{\extracolsep{\fill}}r}
\ifthenelse{\boolean{pdflatex}}
{\vspace*{-2.7cm}\mbox{\!\!\!\includegraphics[width=.14\textwidth]{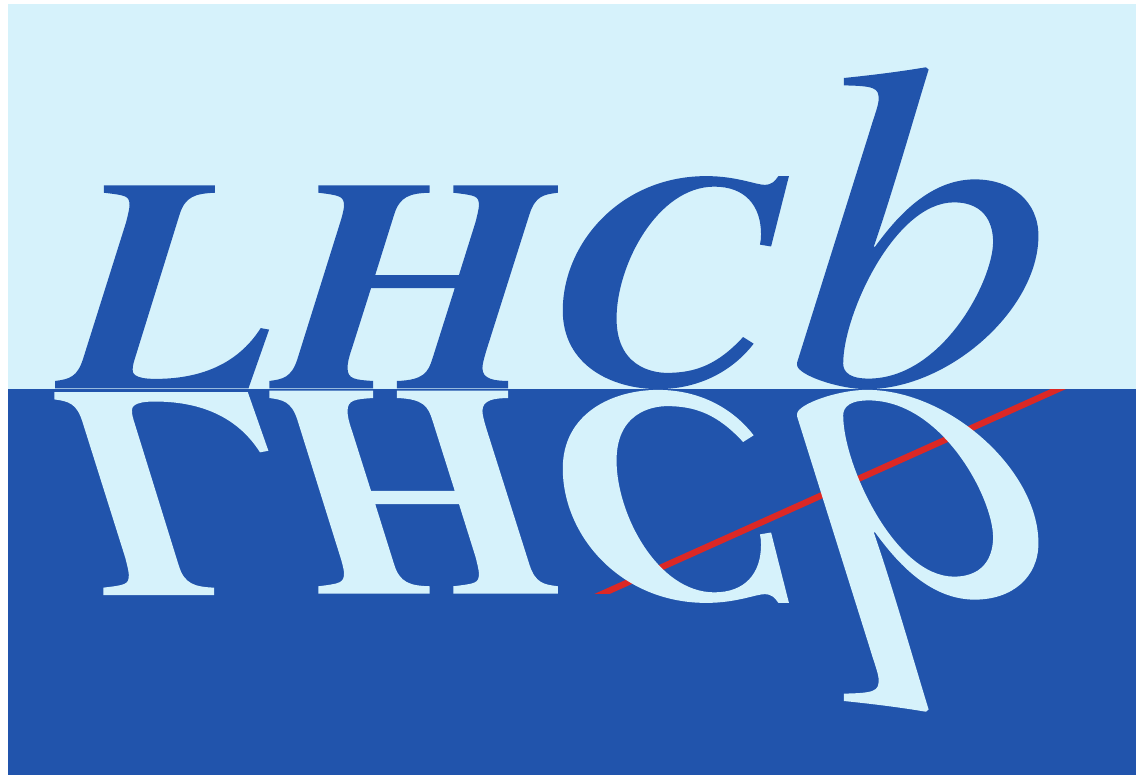}} & &}%
{\vspace*{-1.2cm}\mbox{\!\!\!\includegraphics[width=.12\textwidth]{lhcb-logo.eps}} & &}%
\\
 & & CERN-PH-EP-2014-173 \\  
 & & LHCb-PAPER-2014-034 \\  
 & & 30 April 2025 \\ 
 & & \\
\end{tabular*}

\vspace*{2.0cm}

{\bf\boldmath\huge
\begin{center}
  Evidence for \CP violation in \ppkplus decays
\end{center}
}

\vspace*{1.0cm}

\begin{center}
The LHCb collaboration\footnote{Authors are listed at the end of this article.}
\end{center}

\vspace{\fill}

\begin{abstract}
  \noindent
Three-body \ppkplus and \pppiplus decays are studied using a data sample corresponding to an integrated luminosity of 3.0\invfb collected by the \lhcb experiment in proton-proton collisions at center-of-mass energies of $7$ and $8$ TeV. Evidence of \CP violation in the \ppkplus decay is found in regions of the phase space, representing the first measurement of this kind for a final state containing baryons. Measurements of the forward-backward asymmetry of the light meson in the $p\antiproton$ rest frame yield $\AFB(p\antiproton \Kp,~m_{p\antiproton}<2.85\gevcc)=0.495\pm0.012~(\mathrm{stat})\pm0.007~(\mathrm{syst})$ and $\AFB(p\antiproton \pip,~m_{p\antiproton}<2.85\gevcc)=-0.409\pm0.033~(\mathrm{stat})\pm0.006~(\mathrm{syst})$. In addition, the branching fraction of the decay $B^+\to\Lbar(1520)p$ is measured to be $\mathcal{B}(\lambdaFifteenTwentypplus)=(3.15\pm0.48~(\mathrm{stat})\pm0.07~(\mathrm{syst})\pm0.26~(\mathrm{BF}))\times 10^{-7}$, where BF denotes the uncertainty on secondary branching fractions.
   
\end{abstract}

\vspace*{1.0cm}

\begin{center}
  Published in:\\ 
  Phys. Rev. Lett. {\bf 113}, 141801 (2014), Erratum: Phys. Rev. Lett. {\bf 134}, 179901 (2025)
\end{center}

\vspace{\fill}

{\footnotesize 
\centerline{\copyright~CERN on behalf of the \lhcb collaboration, license \href{http://creativecommons.org/licenses/by/4.0/}{CC-BY-4.0}.}}
\vspace*{2mm}

\end{titlepage}


\newpage
\setcounter{page}{2}
\mbox{~}


\cleardoublepage

\renewcommand{\thefootnote}{\arabic{footnote}}
\setcounter{footnote}{0}


\pagestyle{plain} 
\setcounter{page}{1}
\pagenumbering{arabic}

Direct \CP violation can appear as a rate asymmetry in the decay of a particle and its \CP conjugate, and can be observed when at least two amplitudes, carrying different weak and strong phases, contribute to the final state. For $B$ mesons, it was observed for the first time in two-body $B^0\to K^+\pi^-$ decays\cite{babar_directCP,belle_directCP}. The weak phases are sensitive to physics beyond the Standard Model that may appear at a high energy scale, and their extraction requires a determination of the relative strong phases. Three-body decays are an excellent laboratory for studying strong phases of interfering amplitudes. In particular, charmless decays of $B^+$ mesons, \kpipiplus, \kkkplus, \pipipiplus, \kkpiplus have been investigated recently \cite{LHCB-PAPER-2013-027,LHCB-PAPER-2013-051,LHCb-PAPER-2014-044}\footnote{Throughout the Letter, the inclusion of charge conjugate processes is implied, except in the definition of \CP asymmetries.}. Similar studies have been conducted for the baryonic final states \ppkplus and \pppiplus\cite{LHCB-PAPER-2013-031}.
In the $B^+\to h^+h^-h^+$ decays ($h=\pi$ or $K$ throughout this Letter), large asymmetries, not necessarily associated to resonances, have been observed in the low $K^+K^-$ and $\pi^+\pi^-$ mass regions. These observations suggest that rescattering between $\pi^+\pi^-$ and $K^+K^-$ pairs may play an important role in the generation of the strong phase difference needed for \CP violation to occur \cite{Wolfenstein}. The \pphplus decays, although sharing the same quark-level diagrams, may exhibit different behavior due to the baryonic nature of two out of the three final-state particles. 

This Letter reports the first evidence for \CP violation in charmless \ppkplus decays. These decays are studied in the region with invariant mass $m_{p\antiproton}<2.85\gevcc$, below the charmonium resonances threshold. In addition, a more accurate measurement of the branching fraction of the decay \lambdaFifteenTwentypplus is performed, using the reconstruction of $\Lbar(1520)\to K^+\antiproton$ decays, and improved determinations of the spectra and angular asymmetries are also reported. The mode $\Bp \to \jpsi(\to p\antiproton)\Kp$ serves as a control channel. The data used have been collected with the \lhcb detector and correspond to 1.0 and 2.0 \invfb of integrated luminosity at $7$ and $8$ TeV center-of-mass energies in $pp$ collisions, respectively. The data samples are analyzed separately and the results are averaged.

The \lhcb detector is a single-arm forward spectrometer covering the \mbox{pseudorapidity} range $2<\eta <5$, described in detail in Ref.~\cite{Alves:2008td}. The detector allows for the reconstruction of both charged and neutral particles. For this analysis, the ring-imaging Cherenkov (RICH) detectors \cite{RichPerf}, distinguishing pions, kaons and protons, are particularly important. 

The analysis uses simulated events generated by \pythia~8.1~\cite{Sjostrand:2008} with a specific \lhcb configuration~\cite{LHCb-PROC-2010-056}.  Decays of hadronic particles are described by \evtgen~\cite{Lange:2001uf} in which final state radiation is generated using \photos~\cite{Golonka:2005pn}. The interaction of the generated particles with the detector and its response are implemented using the \geant toolkit~\cite{Allison:2006ve,*Agostinelli:2002hh} as described in Ref.~\cite{LHCb-PROC-2011-006}. Nonresonant \pphplus events are simulated, uniformly distributed in phase space, to study the variation of efficiencies across the Dalitz \cite{Dalitz} plane, as well as resonant modes such as $\Bp \to \jpsi(\to p\antiproton)\Kp$, $\Bp \to \etac(\to p\antiproton)\Kp$, $\Bp \to \psitwos(\to p\antiproton)\Kp$, $\Bp \to \Lbar(1520)(\to K^+\antiproton)p$, and $\Bp \to \jpsi(\to p\antiproton)\pip$.

Three charged particles are combined to form \pphplus decay candidates. The discrimination of signal from background is done through a multivariate analysis using a boosted decision tree (BDT) classifier\cite{BDT_theory}. Input quantities include kinematical and topological variables related to the $\Bp$ candidates and the individual tracks. The momentum, vertex and flight distance of the $\Bp$ candidate are exploited, and track fit quality criteria, impact parameter and momentum information of final-state particles are also used. The BDT is trained using simulated signal events, and events in the high sideband of the $p\bar{p}h^+$ invariant mass ($5.4<m(p\bar{p}h^+)<5.5~\gevcc$), which represent the background. 
Tight particle identification (PID) requirements are applied to reduce the combinatorial background and suppress the cross-feed between $p\antiproton \Kp$ and $p\antiproton \pip$. The PID efficiencies are derived from calibration data samples of kinematically identified pions, kaons and protons originating from the decays $D^{*+}\to D^0(\to K^-\pi^+)\pi^+$ and $\Lz\to p\pi^-$.

Signal and background yields are extracted using unbinned extended maximum likelihood fits to the invariant mass distribution of the $p\antiproton h^+$ combinations.
The \ppkplus signal is modeled by the sum of two Crystal Ball functions \cite{CB_shape}, for which the common mean and core width are allowed to float in the fit. Beside the signal component, the fit includes the parameterizations of the combinatorial background and partially reconstructed $B\to p\antiproton K^*$ decays, where a pion from the $K^*$ decay is not reconstructed, resulting in a $p\antiproton K$ invariant mass below the nominal $B$ mass. An asymmetric Gaussian function with power-law tails is used to model a possible $p\antiproton \pip$ cross-feed component, where the pion is misidentified as a kaon. This contribution is found to be small.

The fit to the \pppiplus decay uses similar parameterizations for the signal, combinatorial background, $p\antiproton \Kp$ cross-feed and partially reconstructed background from $B\to p\antiproton \rho$ decays (with a missing pion from the $\rho$ decay). The cross-feed is found to be negligible.

\begin{figure*}[t]
\begin{center}
\includegraphics[width=0.45\textwidth]{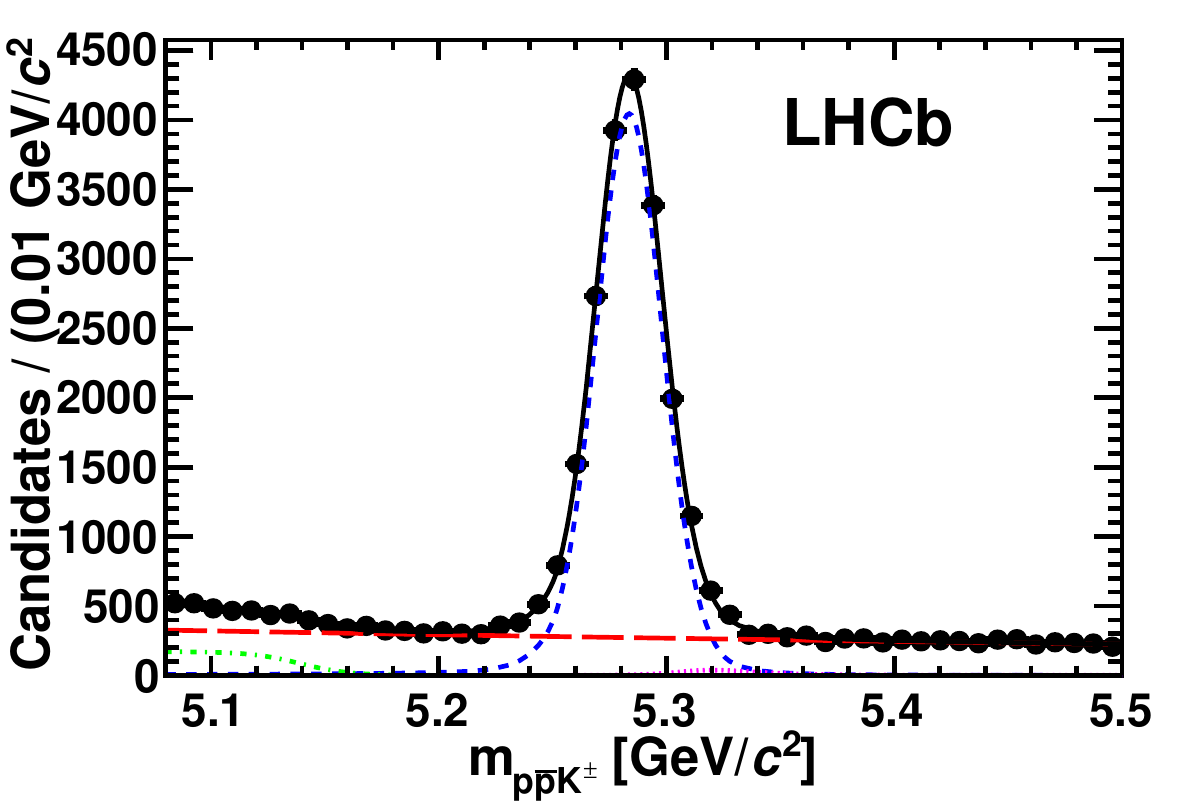}
\includegraphics[width=0.45\textwidth]{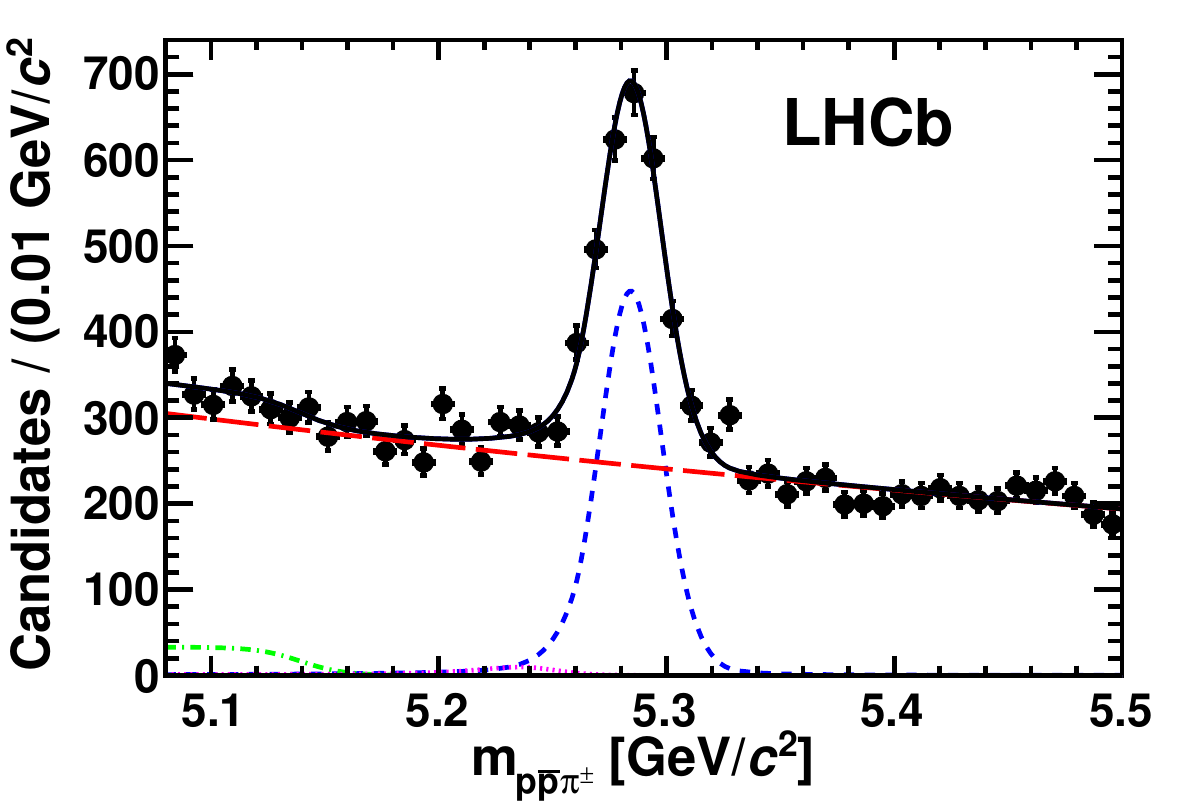}
\end{center}
\caption{Invariant mass distributions of (left) $p\antiproton \Kp$ and (right) $p\antiproton \pip$ candidates. The points with error bars represent data. The solid black line represents the total fit function. The components are represented by blue dashed (signal), purple dotted (cross-feed), red long-dashed (combinatorial background) and green dashed-dotted (partially reconstructed background) curves.}
\label{Fig:pph_globalfits}
\end{figure*}

The \pphplus invariant mass spectra are shown in Fig.~\ref{Fig:pph_globalfits}.
The signal yields obtained from the fits are $N(p\antiproton K^\pm)=18\,721\pm142$ and $N(p\antiproton \pi^\pm)=1988\pm74$, where the uncertainties are statistical only.

\begin{figure*}[t]
\begin{center}
\includegraphics[width=0.45\textwidth]{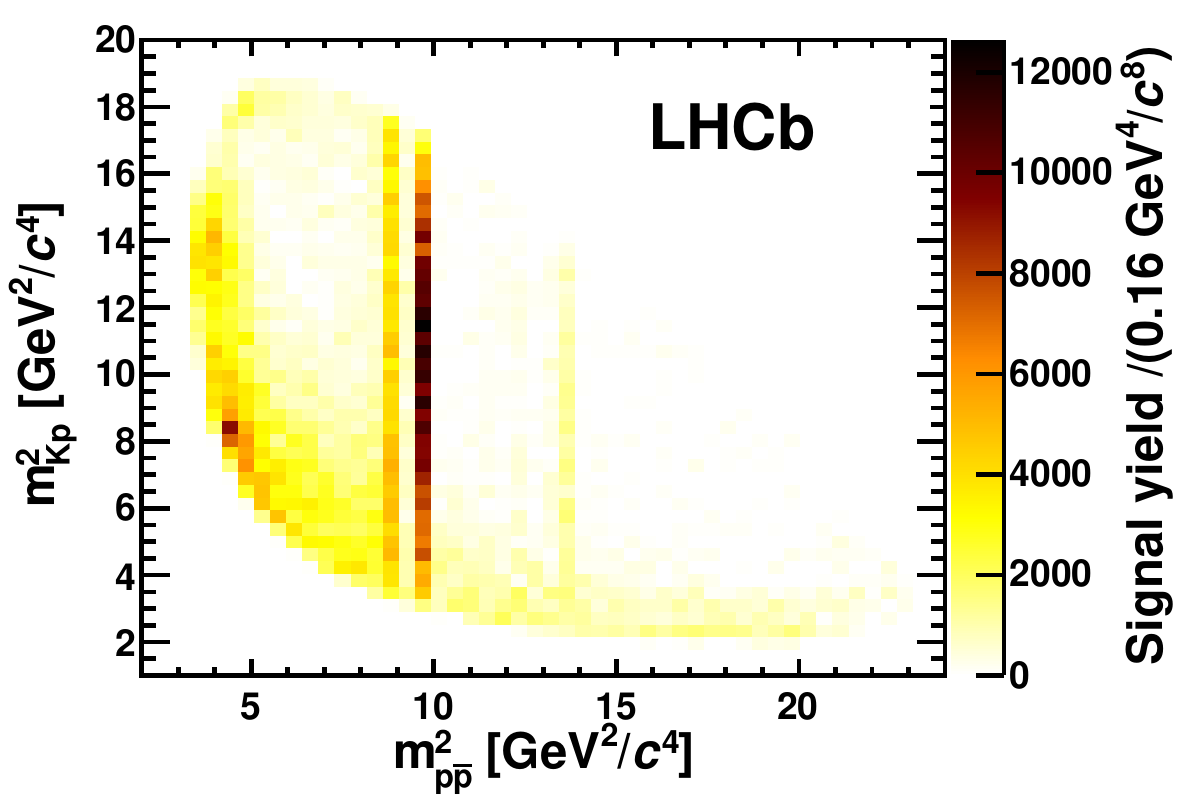}
\includegraphics[width=0.45\textwidth]{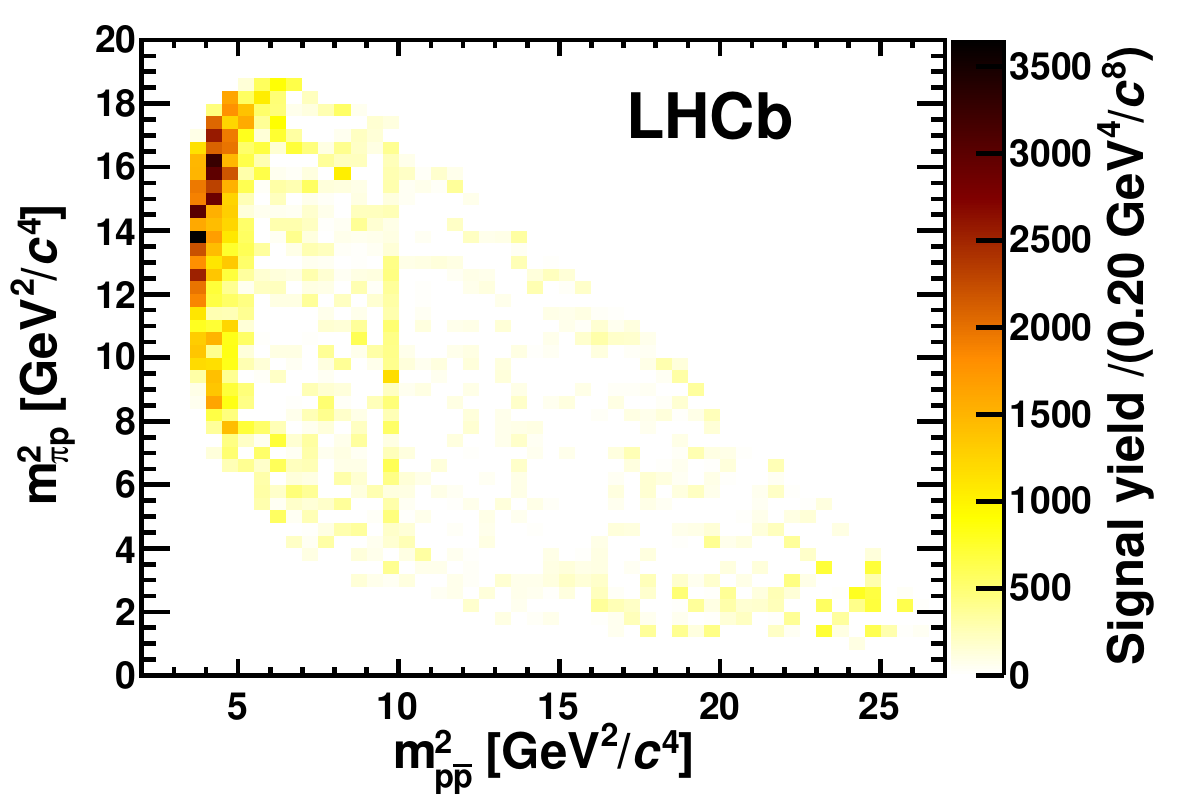}
\end{center}
\caption{Background-subtracted and acceptance-corrected Dalitz-plot distributions for (left) \ppkplus and (right) \pppiplus.}
\label{Fig:pph_dalitz_splots}
\end{figure*}

\begin{figure}[t]
\begin{center}
\includegraphics[width=0.45\textwidth]{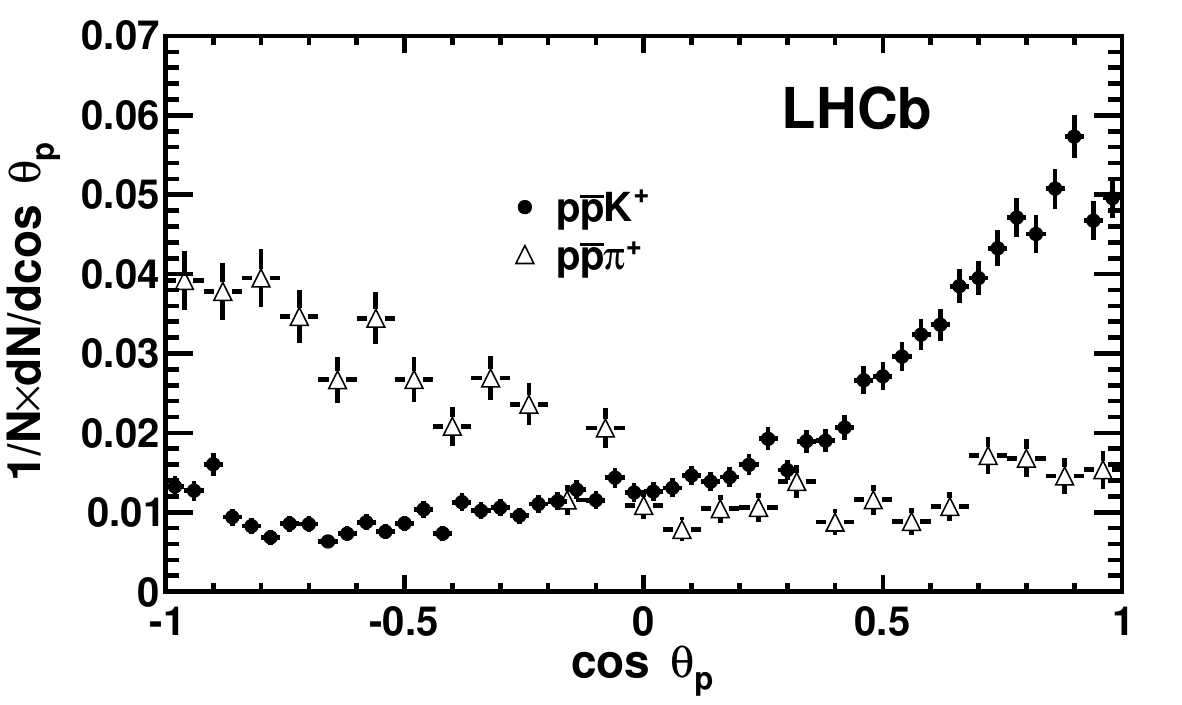}
\end{center}
\caption{Background-subtracted and acceptance-corrected normalized distributions of $\cos\theta_p$ for \ppkplus and \pppiplus decays, for $m_{p\antiproton}<2.85\gevcc$. The data points are shown with their total uncertainties.}
\label{Fig:pph_costhetadiffspectra}
\end{figure}

The distribution of events in the Dalitz plane, defined by $(m_{p\antiproton}^2,m_{hp}^2)$ where $hp$ denotes the neutral combinations $h^-p$ and $h^+\antiproton$, is examined. From the fits to the $B^+$ candidate invariant mass, shown in Fig.~\ref{Fig:pph_globalfits}, signal weights are calculated with the {\it sPlot} technique \cite{sPlots_NIM}. These weights are corrected for trigger, reconstruction and selection efficiencies, which are estimated with simulated samples and calibration data. The Dalitz-plot variables are calculated constraining the $\proton\antiproton h^+$ invariant mass to the known $B^+$ meson mass \cite{DTF,PDG}. 
Figure \ref{Fig:pph_dalitz_splots} shows the Dalitz distributions of the \pphplus events.
Similarly to the results reported in Ref.\cite{LHCB-PAPER-2013-031,LHCb-PAPER-2012-047}, clear signals of \jpsi, \etac and \psitwos resonances are observed, while \ppkplus and \pppiplus non-charmonium events both accumulate near the $p\antiproton$ threshold. However, \ppkplus events preferentially occupy the region with low $Kp$ invariant mass while \pppiplus events populate the region with large $\pi p$ invariant mass. 
This difference in the Dalitz distribution can also be observed as a difference in the distribution of the helicity angle $\theta_p$ of the $p\antiproton $ system, defined as the angle between the charged meson $h$ and the oppositely charged baryon in the rest frame of the $p\antiproton $ system. The distributions of $\cos(\theta_p)$ are depicted in Fig.~\ref{Fig:pph_costhetadiffspectra}.
 
Data and simulation are used to assign systematic uncertainties, accounting for the PID correction and fit model, to the angular and charge asymmetries, and to the relative branching fractions. The systematic uncertainty associated to the PID correction cancels in the asymmetry measurements.

\begin{figure}[t]
\begin{center}
\includegraphics[width=0.45\textwidth]{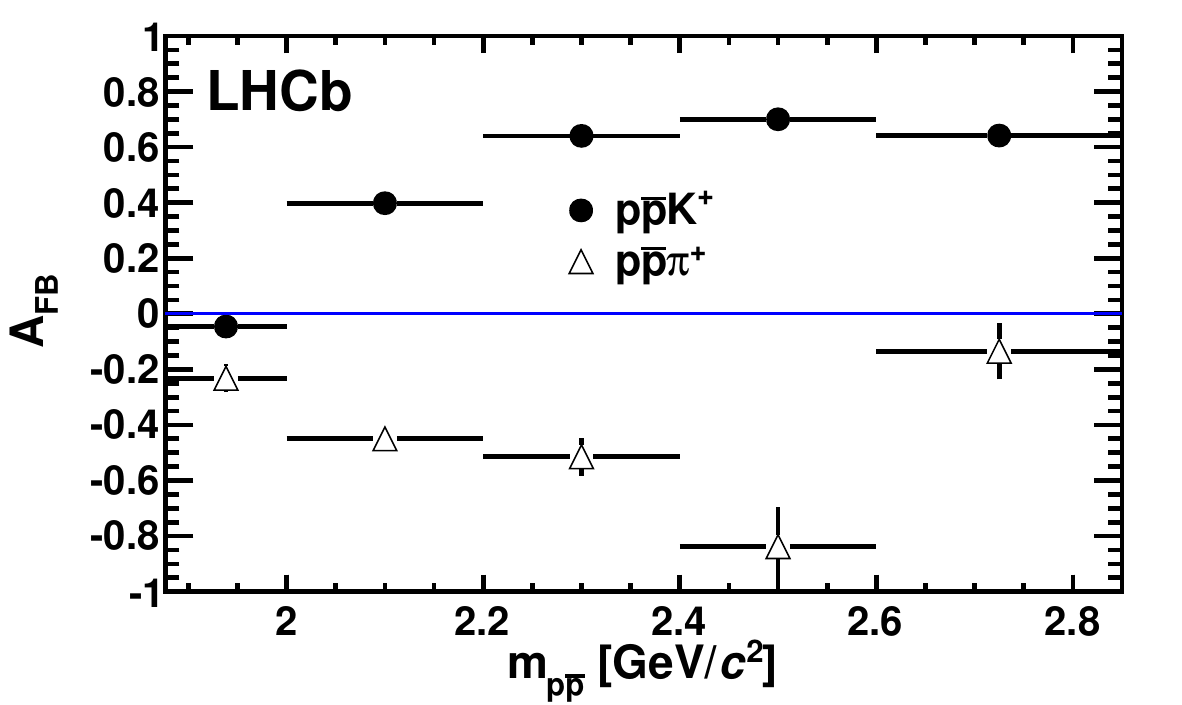}
\end{center}
\caption{Forward-backward asymmetry in bins of $m_{p\antiproton}$ for \ppkplus and \pppiplus decays. The data points are shown with their total uncertainties.}
\label{Fig:pph_theta_asym}
\end{figure}

The forward-backward asymmetry is measured as
\begin{equation}
\AFB=\frac{N_{\mathrm{pos}}-N_{\mathrm{neg}}}{N_{\mathrm{pos}}+N_{\mathrm{neg}}},
\end{equation}
where $N_{\mathrm{pos}}$ ($N_{\mathrm{neg}}$) is the efficiency-corrected yield for $\cos\theta_p>0$ ($\cos\theta_p<0$). The obtained asymmetries are $\AFB(p\antiproton \Kp,~m_{p\antiproton}<2.85\gevcc)=0.495\pm0.012~(\mathrm{stat})\pm0.007~(\mathrm{syst})$ and $\AFB(p\antiproton \pip,~m_{p\antiproton}<2.85\gevcc)=-0.409\pm0.033~(\mathrm{stat})\pm0.006~(\mathrm{syst})$, where the systematic uncertainty is due to the ratio of average efficiencies in the regions $\cos\theta_p>0$ and $\cos\theta_p<0$. As reported in previous studies \cite{LHCB-PAPER-2013-031,Belle_pph}, the value for \ppkplus contradicts the short-range analysis expectation \cite{suzuki}.
The values of \AFB in bins of $m_{p\antiproton}$ are shown in Fig.~\ref{Fig:pph_theta_asym}; in both cases, they depend strongly on $m_{p\antiproton}$.

The yields of the decays \pphplus in the region $m_{p\antiproton}<2.85\gevcc$ are obtained in the same way as for the integrated signals. Those of the resonant modes are extracted through two-dimensional extended unbinned maximum likelihood fits to invariant mass distributions of $p\antiproton h^+$ and $p\antiproton$ or $K^+\antiproton$, using the same signal and background models for $m_{p\antiproton}$ or $m_{K^+\antiproton}$ as in Ref.~\cite{LHCB-PAPER-2013-031}. The results are shown in Table \ref{Tab:pph_resonant_yields}.
\begin{table*}[tb]
\caption{ Event yields and selection efficiency for \ppkplus and \pppiplus final states.}
  \begin{center}
\begin{tabular}{l r@{$\pm$}l c c}\hline\hline
\centering
 Mode & \multicolumn{2}{c}{Yield} & \multicolumn{1}{c}{Efficiency (\%)}\\ 
\hline 
$\Bp\to \jpsi(\to p\antiproton ) \Kp$ & 4260&67 & 1.55$\pm$0.02\\
$\Bp\to \etac(\to p\antiproton ) \Kp$  & 2182&64 & 1.47$\pm$0.02\\
$\Bp\to \psitwos(\to p\antiproton ) \Kp$ & 368&20 & 1.59$\pm$0.02\\
$\lambdaFifteenTwentypplusdet$ & 128&20 & 1.39$\pm$0.01\\
\ppkplus, $m_{p\antiproton}<2.85\gevcc$ & 8510&104 & 1.58$\pm$0.02\\
\hline
$\Bp\to \jpsi(\to p\antiproton ) \pip$& 122&12 &1.07$\pm$0.01\\
\pppiplus, $m_{p\antiproton}<2.85\gevcc$ & 1632&64 & 1.15$\pm$0.01\\
\hline\hline
\end{tabular}
\end{center}
    \label{Tab:pph_resonant_yields}
\end{table*}
The branching fractions of the decays $\Bp\to\Lbar(1520)(\to K^+\antiproton )p$ and $\pppiplus,~m_{p\antiproton}<2.85\gevcc$ are measured relative to the \jpsi modes as
\begin{eqnarray}
\nonumber \frac{ \mathcal{B}(\Bp\to\Lbar(1520)(\to K^+\antiproton )p) }{\mathcal{B}(\Bp \to \jpsi(\to p\antiproton )\Kp)}&=& 0.033\pm0.005~(\mathrm{stat})\pm 0.007~(\mathrm{syst}),\\
\nonumber \frac{ \mathcal{B}(\pppiplus,~m_{p\antiproton}<2.85\gevcc) }{\mathcal{B}(\Bp \to \jpsi(\to p\antiproton )\pip)}&=& 12.0\pm1.2~(\mathrm{stat})\pm 0.3~(\mathrm{syst}).
\end{eqnarray}
The systematic uncertainties also include contributions from the background model.
Using $\mathcal{B}(\Bp \to \jpsi\Kp)= (1.016\pm0.033)\times 10^{-3}$, $\mathcal{B}(\Bp \to \jpsi\pip)= (4.1\pm0.4)\times 10^{-5}$, $\mathcal{B}(\jpsi\to p\antiproton )=(2.17\pm0.07)\times 10^{-3}$ \cite{PDG}, and $\mathcal{B}(\Lambda(1520)\to K^-p)=0.234\pm0.016$ \cite{Saphir}, the branching fractions are measured to be:
\begin{center}
$\mathcal{B}(\lambdaFifteenTwentypplus)=(3.15\pm0.48~(\mathrm{stat})\pm0.07~(\mathrm{syst})\pm0.26~(\mathrm{BF}))\times 10^{-7}$,\\
$\mathcal{B}(\pppiplus,~m_{p\antiproton}<2.85\gevcc)=(1.07\pm 0.11~(\mathrm{stat})\pm 0.03~(\mathrm{syst})\pm 0.11~(\mathrm{BF}))\times 10^{-6}$,
\end{center}
where BF denotes the uncertainty on the aforementioned secondary branching fractions. The former measurement supersedes what is reported in Ref.~\cite{LHCB-PAPER-2013-031}.

\begin{figure}[t]
\begin{center}
\includegraphics[width=0.5\textwidth]{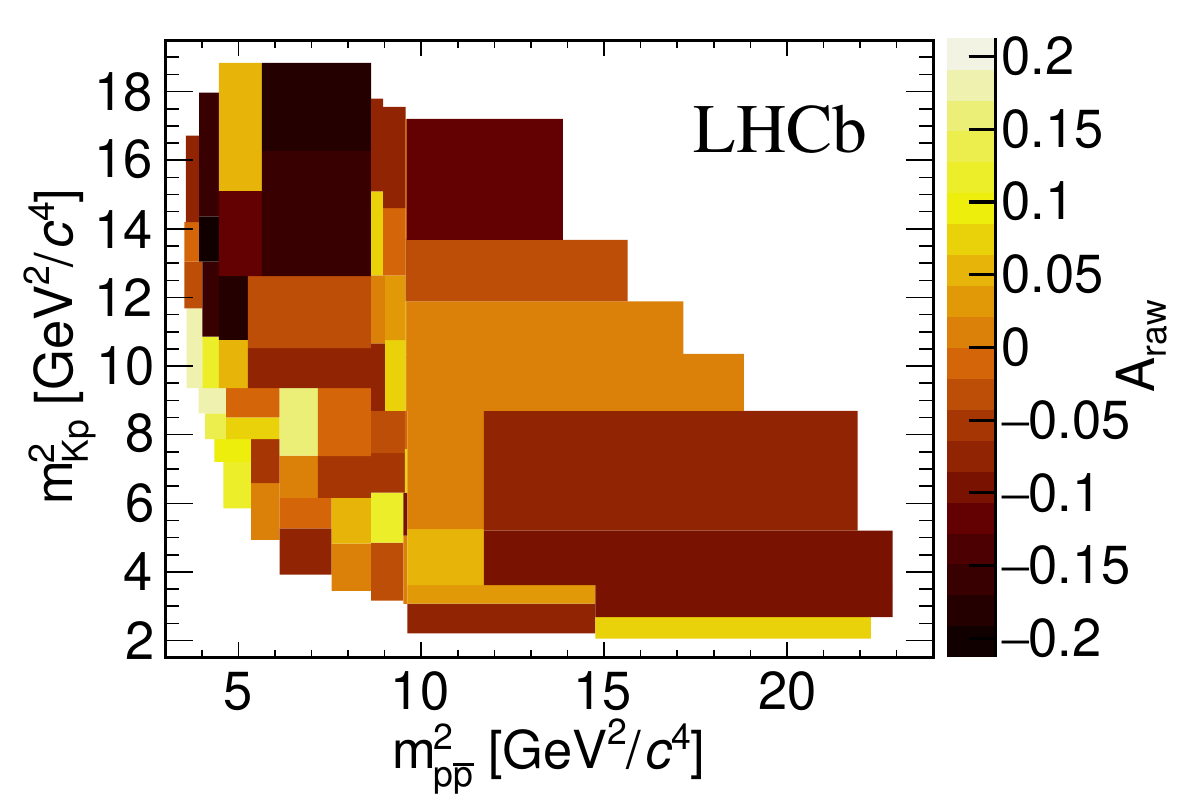}
\end{center}
\caption{Asymmetries of the number of signal events in bins of the Dalitz-plot variables for \ppk. The number of events in each bin is approximately 300.}
\label{Fig:acp_2D_color_ppk}
\end{figure}

The raw charge asymmetry is measured from the yields $N$ as
\begin{equation}
\acpraw = \frac{N(B^-\to p\antiproton h^-)-N(B^+\to p\antiproton h^+)}{N(B^-\to p\antiproton h^-)+N(B^+\to p\antiproton h^+)},
\end{equation}
and is investigated in the Dalitz plane using signal weights inferred from the fits shown in Fig.~\ref{Fig:pph_globalfits}, for \Bm and \Bp samples. This asymmetry includes production and detection asymmetries. The statistics of the \pppi decays is not sufficient to perform such an analysis, so only the \ppk case is studied. An adaptative binning algorithm is used so that the sum of \Bm and \Bp events in each bin is approximately constant. Figure \ref{Fig:acp_2D_color_ppk} shows the distribution of \acpraw in the Dalitz plane.

\begin{figure}[t]
\begin{center}
\includegraphics[width=0.5\textwidth]{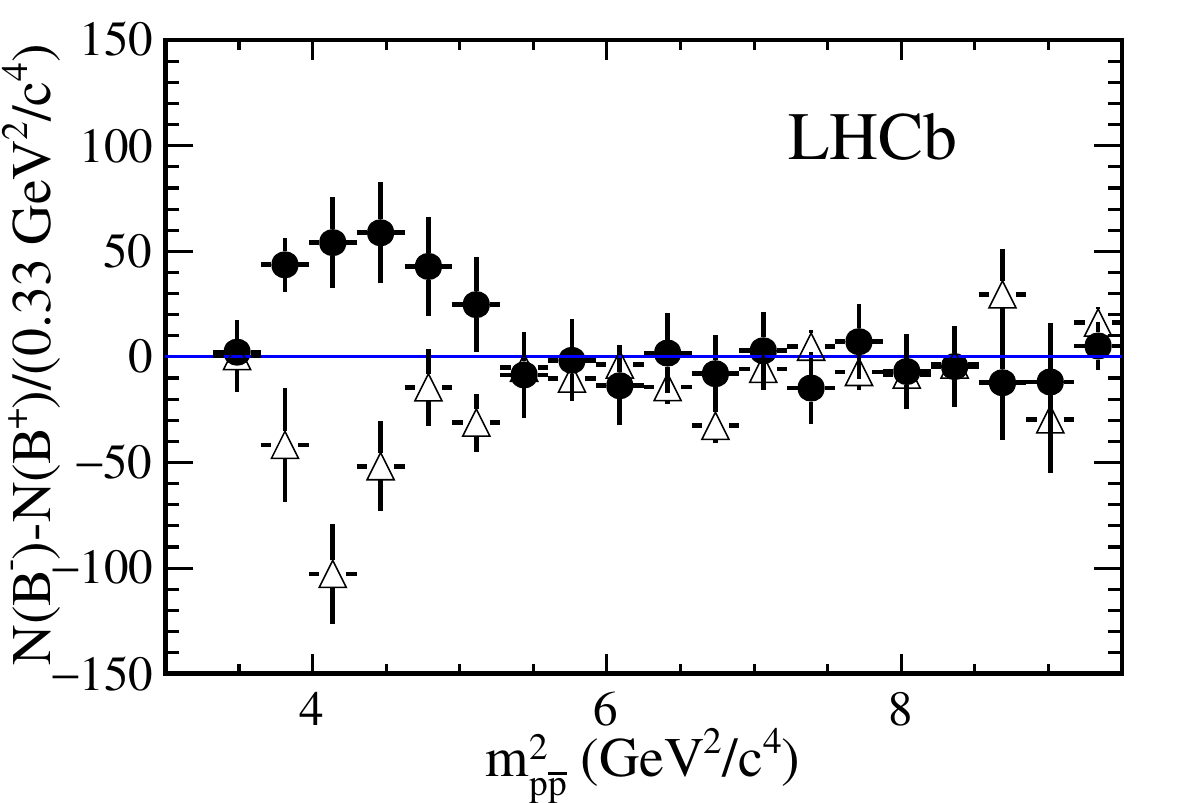}
\end{center}
\caption{$N(B^-)-N(B^+)$ in bins of $m_{p\antiproton}^2$ for $m_{Kp}^2<10\gevgevcccc$ (black filled circles) and $m_{Kp}^2>10\gevgevcccc$ (open triangles).}
\label{Fig:acp_mpp2proj_bdiff_ppk}
\end{figure}

A clear pattern is observed near the $p\antiproton$ threshold where \acpraw is positive for $m_{Kp}^2<10\gevgevcccc$ and negative for $m_{Kp}^2>10\gevgevcccc$. Figure \ref{Fig:acp_mpp2proj_bdiff_ppk} shows the $m_{p\antiproton}^2$ projections of $N(B^-)-N(B^+)$ in the regions of interest.

To quantify the effect, unbinned extended maximum likelihood simultaneous fits to $B^-$ and $B^+$ samples are performed in regions of the Dalitz plane, using the same models as the global fits. 
The raw asymmetry is corrected for acceptance, by taking into account the small difference in average efficiency due to the $B^-$ and $B^+$ samples populating differently the Dalitz plane. Physical asymmetries are obtained after acceptance correction ($\acprawacc$) and accounting for the production $\aprod(\Bpm)$ and kaon detection $\adet(\Kpm)$ asymmetries:
\begin{equation}
\acp = \acprawacc - \aprod(\Bpm)-\adet(\Kpm).
\end{equation}
The decay $\Bpm\to\jpsi(p\antiproton)\Kpm$, part of the selected sample, is used to determine $A_\Delta=\aprod(\Bpm) + \adet(\Kpm)$:
\begin{equation}
A_\Delta = \acpraw( \Bpm\to\jpsi(p\antiproton ) \Kpm)-\acp(\Bpm\to\jpsi\Kpm).
\end{equation}
The value $\acp(\Bpm\to\jpsi\Kpm)=(0.6\pm0.4)\%$ is taken from Ref.~\cite{D0_jpsik}. When using $\acpraw( \Bpm\to\jpsi(p\antiproton ) \Kpm)$, differences in the momentum asymmetry of the $p\antiproton$ pair between $\Bpm\to\jpsi(p\antiproton ) \Kpm$ and nonresonant $\Bpm\to p\antiproton\Kpm$ decays are accounted for.
A similar procedure is applied to obtain $\acp(\Bpm\to\etac(p\antiproton)\Kpm)$ and $\acp(\Bpm\to\psitwos(p\antiproton)\Kpm)$.
The \pppi decays are also considered in the region $m_{p\antiproton}<2.85\gevcc$. In this case, the correction also involves the pion detection asymmetry, $A_\Delta^\prime = \acpraw(\Bpm\to \jpsi(p\antiproton ) \Kpm)-\acp(\Bpm\to\jpsi\Kpm)-\adet(\Kpm)+\adet(\pipm)$. The value $\adet(\Kpm)-\adet(\pipm)=(-1.2\pm0.1)\%$ is taken from studies of prompt $D^+$ decays \cite{LHCb-PAPER-2014-013}.
Table \ref{Tab:acp_ppk} shows the results, including asymmetries of resonant modes. Closer to the $p\antiproton$ threshold enhancement, $m_{p\antiproton}^2<6\gevgevcccc$, the asymmetry is found to reach the value $0.091\pm0.026~(\mathrm{stat})\pm0.004~(\mathrm{syst})$, for $m_{Kp}^2<10\gevgevcccc$.

\begin{table*}[tb]
\caption{\CP asymmetries for \ppk and \pppi decays. The systematic uncertainties are dominated by the precision on the measurement $\acp(\Bpm\to\jpsi\Kpm)$.}
  \begin{center}
\begin{tabular}{l r@{$\pm$}l@{$\pm$}l}\hline\hline
\centering
 Mode/region & \multicolumn{3}{c}{\acp}\\ 
\hline 
$\etac(p\antiproton ) \Kpm$  & $-$0.027&0.034$~(\mathrm{stat})$&0.004$~(\mathrm{syst})$\\
$\psitwos(p\antiproton ) \Kpm$ & $-$0.079&0.059$~(\mathrm{stat})$&0.004$~(\mathrm{syst})$\\
$p\antiproton\Kpm,~m_{p\antiproton}<2.85\gevcc$ & $-$0.006&0.020$~(\mathrm{stat})$&0.004$~(\mathrm{syst})$\\
$p\antiproton\Kpm,~m_{p\antiproton}<2.85\gevcc,~m_{Kp}^2<10\gevgevcccc$ & 0.056&0.023$~(\mathrm{stat})$&0.004$~(\mathrm{syst})$\\
$p\antiproton\Kpm,~m_{p\antiproton}<2.85\gevcc,~m_{Kp}^2>10\gevgevcccc$ & $-$0.082&0.024$~(\mathrm{stat})$&0.004$~(\mathrm{syst})$\\
$p\antiproton\pipm,~m_{p\antiproton}<2.85\gevcc$ & 0.039&0.039$~(\mathrm{stat})$&0.005$~(\mathrm{syst})$\\
\hline\hline
\end{tabular}
\end{center}
    \label{Tab:acp_ppk}
\end{table*}
The systematic uncertainties are estimated by using alternative fit functions and splitting the data sample according to trigger requirements and magnet polarity. The overall systematic uncertainties are dominated by the uncertainty on the $\acp(\Bpm\to\jpsi\Kpm)$ measurement.

In summary, an interesting sign-inversion pattern of the \CP asymmetry appears at low $p\antiproton$ invariant masses in \ppk decays. Although this resembles what is observed at low $h^+h^-$ masses in the \hhh decays, the strong phase difference could involve a specific mechanism such as interfering long-range $p\antiproton$ waves with different angular momenta \cite{suzuki}. In the region $m_{p\antiproton}<2.85\gevcc,~m_{Kp}^2>10\gevgevcccc$, the measured asymmetry is negative with a significance of nearly 3.5$\sigma$, which represents the first evidence of \CP violation in $b$-hadron decays with baryons in the final state. In the region ($m_{p\antiproton}^2<6\gevgevcccc$, $m_{Kp}^2<10\gevgevcccc$), the asymmetry is positive with a significance of 3.5$\sigma$.\\
The $h$ hadron forward-backward asymmetry in non-charmonium \pphplus decays is measured as $\AFB(p\antiproton \Kp,~m_{p\antiproton}<2.85\gevcc)=0.495\pm0.012~(\mathrm{stat})\pm0.007~(\mathrm{syst})$ and $\AFB(p\antiproton \pip,~m_{p\antiproton}<2.85\gevcc)=-0.409\pm0.033~(\mathrm{stat})\pm0.006~(\mathrm{syst})$. These asymmetries could be interpreted as being due to the dominance of nonresonant $p\antiproton$ scattering \cite{suzuki}.
Finally, an improved measurement of $\mathcal{B}(\lambdaFifteenTwentypplus)=(3.15\pm0.48~(\mathrm{stat})\pm0.07~(\mathrm{syst})\pm0.26~(\mathrm{BF}))\times 10^{-7}$ is obtained.

\section*{Acknowledgements}

\noindent We express our gratitude to our colleagues in the CERN
accelerator departments for the excellent performance of the LHC. We
thank the technical and administrative staff at the LHCb
institutes. We acknowledge support from CERN and from the national
agencies: CAPES, CNPq, FAPERJ and FINEP (Brazil); NSFC (China);
CNRS/IN2P3 (France); BMBF, DFG, HGF and MPG (Germany); SFI (Ireland); INFN (Italy); 
FOM and NWO (The Netherlands); MNiSW and NCN (Poland); MEN/IFA (Romania); 
MinES and FANO (Russia); MinECo (Spain); SNSF and SER (Switzerland); 
NASU (Ukraine); STFC (United Kingdom); NSF (USA).
The Tier1 computing centres are supported by IN2P3 (France), KIT and BMBF 
(Germany), INFN (Italy), NWO and SURF (The Netherlands), PIC (Spain), GridPP 
(United Kingdom).
We are indebted to the communities behind the multiple open 
source software packages on which we depend. We are also thankful for the 
computing resources and the access to software R\&D tools provided by Yandex LLC (Russia).
Individual groups or members have received support from 
EPLANET, Marie Sk\l{}odowska-Curie Actions and ERC (European Union), 
Conseil g\'{e}n\'{e}ral de Haute-Savoie, Labex ENIGMASS and OCEVU, 
R\'{e}gion Auvergne (France), RFBR (Russia), XuntaGal and GENCAT (Spain), Royal Society and Royal
Commission for the Exhibition of 1851 (United Kingdom).

\ifx\mcitethebibliography\mciteundefinedmacro
\PackageError{LHCb.bst}{mciteplus.sty has not been loaded}
{This bibstyle requires the use of the mciteplus package.}\fi
\providecommand{\href}[2]{#2}

\newpage
\centerline{\large\bf LHCb collaboration}
\begin{flushleft}
\small
R.~Aaij$^{41}$, 
B.~Adeva$^{37}$, 
M.~Adinolfi$^{46}$, 
A.~Affolder$^{52}$, 
Z.~Ajaltouni$^{5}$, 
S.~Akar$^{6}$, 
J.~Albrecht$^{9}$, 
F.~Alessio$^{38}$, 
M.~Alexander$^{51}$, 
S.~Ali$^{41}$
% ...

{\footnotesize \it
% ...
$ ^{5}$Clermont Universit\'{e}, Universit\'{e} Blaise Pascal, CNRS/IN2P3, LPC, Clermont-Ferrand, France\\
$ ^{6}$CPPM, Aix-Marseille Universit\'{e}, CNRS/IN2P3, Marseille, France\\
% ...
$ ^{9}$Fakult\"{a}t Physik, Technische Universit\"{a}t Dortmund, Dortmund, Germany\\
% ...
$ ^{37}$Universidad de Santiago de Compostela, Santiago de Compostela, Spain\\
$ ^{38}$European Organization for Nuclear Research (CERN), Geneva, Switzerland\\
% ...
$ ^{41}$Nikhef National Institute for Subatomic Physics, Amsterdam, The Netherlands\\
% ...
$ ^{46}$H.H. Wills Physics Laboratory, University of Bristol, Bristol, United Kingdom\\
% ...
$ ^{51}$School of Physics and Astronomy, University of Glasgow, Glasgow, United Kingdom\\
$ ^{52}$Oliver Lodge Laboratory, University of Liverpool, Liverpool, United Kingdom\\
% ...
}
\end{flushleft}

\end{document}